\documentclass[fleqn,twoside]{article}
\usepackage{espcrc2}
\usepackage{graphicx}
\usepackage[figuresright]{rotating}

\newcommand{\AmS}{{\protect\the\textfont2
A\kern-.1667em\lower.5ex\hbox{M}\kern-.125emS}}

\newcommand{\like}{L}
\newcommand{\mdl}{\mathcal{M}}
\title{Density perturbations from both the
inflaton and the curvaton}

\author{George Lazarides\address{Physics
Division, School of Technology, Aristotle
University of Thessaloniki, \\ Thessaloniki
54124, Greece}%
\thanks{lazaride@eng.auth.gr}}

\begin{document}

\begin{abstract}
We consider a supersymmetric grand unified
model which leads to hybrid inflation and
solves the strong CP and $\mu$ problems via a
Peccei-Quinn symmetry, with the Peccei-Quinn
field acting as a curvaton generating together
with the inflaton the curvature perturbation.
The model yields an isocurvature perturbation
too of mixed correlation with the adiabatic
one. Two choices of
parameters are confronted with the Wilkinson
microwave anisotropy probe and other cosmic
microwave background radiation data. For the
choice giving the best fitting, the curvaton
contribution to the amplitude of the adiabatic
perturbation must be smaller than $67\%$ at
$95\%$ confidence level and the best-fit power
spectra are dominated by the adiabatic inflaton
contribution. This case is disfavored relative
to the pure inflaton scale-invariant case with
odds of 50 to 1. For the second choice, the
adiabatic mode is dominated by the curvaton,
but this case is strongly disfavored relative
to the pure inflaton case (with odds of $10^7$
to 1). Thus, in this model, the perturbations
must be dominated by the adiabatic component
from the inflaton.
\vspace{1pc}
\end{abstract}

\maketitle

\section{INTRODUCTION}

\par
The usual assumption \cite{llbook,lectures} in
inflationary cosmology is that the
density perturbation is generated solely by the
slowly rolling inflaton field and is, thus,
expected to be purely adiabatic and Gaussian.
However, lately, the alternative possibility
\cite{curv1,curv2} that the adiabatic density
perturbations originate from the inflationary
perturbations of some other ``curvaton'' field
which is also light during inflation has
attracted much attention. In this case,
appreciable isocurvature perturbations
\cite{curv2,curv3} in the densities of the
various components of the cosmic fluid as well
as significant non-Gaussianity can arise. The
main reason for advocating this curvaton
hypothesis is that it makes \cite{dimo} the
task of constructing viable inflationary models
much easier, since it liberates us from the
restrictive requirement that the inflaton is
responsible for the curvature perturbation.

\par
The standard curvaton hypothesis
\cite{curv1,curv2}, which insists that the
total curvature perturbation originates
solely from the curvaton, can also be quite
restrictive and not so natural. Indeed, in this
case, one needs to impose \cite{dimo} an upper
bound on the inflationary scale in order to
ensure that the perturbation from the inflaton
is negligible. This bound can be quite strong
if the slow-roll parameter $\epsilon$ is small.
Moreover, in generic models, one expects that
all the scalars which are essentially massless
during inflation contribute to the total
curvature perturbation. So, in the presence of
a curvaton, it is natural to assume that
the adiabatic density perturbation is partly
due to this field and partly to the inflaton.
Finally, the recent measurements on the cosmic
microwave background radiation (CMBR) by the
Wilkinson microwave anisotropy probe (WMAP)
satellite \cite{wmap1} have considerably
strengthened \cite{wmap2,iso1,gm} the bound on
the isocurvature perturbation. Thus, the
viability of many curvaton models is in doubt.

\par
We, thus, take \cite{iso} a more liberal and
natural attitude allowing a significant part of
the total curvature perturbation to originate
from the inflaton. We can then hopefully relax
the tension between curvaton models and the
WMAP data without losing the main advantage of
the curvaton hypothesis, which is that it
facilitate the construction of viable
inflationary models.

\section{THE PARTICLE PHYSICS MODEL}

In order to explore this possibility, we take
\cite{iso} the concrete supersymmetric (SUSY)
grand unified theory (GUT) model of
Ref.~\cite{hier}, which is based on the
left-right symmetric
gauge group $G_{LR}={\rm SU}(3)_c\times
{\rm SU}(2)_L\times{\rm SU}(2)_R\times
{\rm U}(1)_{B-L}$. This model leads \cite{hier}
naturally to standard SUSY hybrid inflation
\cite{lyth,dss}, which is the most promising
inflationary scenario. It also solves
simultaneously the strong CP and $\mu$ problems
via a Peccei-Quinn (PQ) symmetry. The PQ field,
which breaks spontaneously the PQ symmetry,
can act as curvaton (see Ref.~\cite{dllr1}).

\par
The ${\rm SU}(2)_{L}$ doublet left-handed quark
and lepton superfields are denoted by $q_i$ and
$l_i$ respectively, whereas the
${\rm SU}(2)_{R}$ doublet antiquark and
antilepton superfields by $q_i^c$ and $l^c_i$
respectively ($i$=1,2,3 is the family index).
The electroweak Higgs superfield $h$ belongs to
a bidoublet $(1,2,2)_0$ representation of
$G_{LR}$. The breaking of $G_{LR}$ to the
standard model gauge group $G_{\rm SM}$,
at a superheavy scale $M\sim 10^{16}~
{\rm{GeV}}$, is achieved through the
superpotential
\begin{equation}
\delta W_1=\kappa S(l_H^c\bar l_H^{c}-M^2),
\label{W1}
\end{equation}
where $l_H^c$, $\bar l_H^{c}$ is a conjugate
pair of ${\rm SU}(2)_R$ doublet left-handed
Higgs superfields with $B-L$ charges equal to
$1,-1$ respectively, and $S$ is a gauge
singlet left-handed superfield. The
dimensionless coupling constant $\kappa$ and
the mass parameter $M$ are made real and
positive by rephasing the fields. The SUSY
minima of the scalar potential lie on the
D-flat direction $l_H^c=\bar l_H^{c*}$ at
$\langle S\rangle=0$, $|\langle l_H^c\rangle|
=|\langle\bar l_H^{c}\rangle|=M$. The model
also contains two extra gauge singlet
left-handed superfields $N$ and $\bar{N}$ for
solving \cite{rsym} the $\mu$ problem via a
PQ symmetry, which also solves the strong
CP problem. They have the following
superpotential couplings:
\begin{equation}
\delta W_2=\frac{\lambda N^2\bar{N}^2}{2m_P}+
\frac{\beta N^2h^2}{2m_P},
\label{W2}
\end{equation}
where $\lambda$ and $\beta$ are dimensionless
coupling constants (made real and positive) and
$m_{\rm P}\simeq 2.44\times 10^{18}~{\rm GeV}$
is the reduced Planck mass.

\par
The model possesses three global ${\rm U}(1)$
symmetries, namely an anomalous PQ symmetry
${\rm U}(1)_{\rm PQ}$, a non-anomalous R
symmetry ${\rm U}(1)_R$, and the baryon number
symmetry ${\rm U}(1)_B$. The PQ and R charges
of the various superfields are
\begin{eqnarray}
PQ:~q^c,l^c,S,l_H^c,\bar l_H^c(0),h,\bar N(1),q,l,
N(-1); \nonumber \\
R:~h,l_H^c,\bar l_H^c,\bar{N}(0),q,q^c,l,l^c,N(1/2),
S(1).
\label{charges}
\end{eqnarray}

\par
The superpotential in Eq.~(\ref{W1}) leads
\cite{lyth,dss} naturally to the standard SUSY
realization of hybrid inflation \cite{hybrid}.
The inflationary path is a built-in classically
flat valley of minima at $l_H^c=\bar l_H^c=0$
and for $|S|$ greater than a critical
(instability) value $S_c=M$. The constant
tree-level potential energy density
$\kappa^{2}M^{4}$ on this path can cause
inflation as well as SUSY breaking leading
\cite{dss} to one-loop radiative corrections
which provide a logarithmic slope along this
path necessary for driving the system towards
the vacua.

\par
The scalar potential $V_{\rm PQ}$ for the PQ
symmetry breaking is derived from the first
term in the right hand side (RHS) of
Eq.~(\ref{W2}) and, after soft SUSY breaking
mediated by minimal supergravity, is
\cite{rsym}
\begin{equation}
\frac{1}{2}m_{3/2}^2\phi^2\left(
1-\frac{|A|\lambda\phi^2}{8m_{3/2}m_P}+
\frac{\lambda^2\phi^4}{16m_{3/2}^2m_P^2}\right),
\label{PQ-pot}
\end{equation}
where $m_{3/2}\sim 1~{\rm TeV}$ is the gravitino
mass and $A$ is the dimensionless coefficient of
the soft SUSY breaking term corresponding to the
first term in the RHS of Eq.~(\ref{W2}). Here,
the phases of $A$, $N$ and $\bar{N}$ are
adjusted and $\vert N\vert$, $\vert\bar{N}\vert$
are taken equal so that $V_{\rm PQ}$ is
minimized. Moreover, rotating $N$ on the real
axis by a R transformation, we defined the
canonically normalized real scalar PQ field
$\phi=2N$. For $|A|>4$, $V_{\rm PQ}$ has a local
minimum at $\phi=0$ and absolute minima at
\begin{equation}
\langle\phi\rangle^2\equiv f_a^2=\frac{2}
{3\lambda}\left(|A|+\sqrt{|A|^2-12}\right)
m_{3/2}m_{\rm P}
\label{eq:fa}
\end{equation}
with $f_a$ being the axion decay constant.
Substituting this vacuum expectation value
(VEV) into the second term in the RHS of
Eq.~(\ref{W2}), we obtain a $\mu$ term with
\begin{equation}
\mu=\frac{\beta f_a^2}{4m_{\rm P}}\sim m_{3/2},
\label{mu}
\end{equation}
as desired \cite{kn}. Note that the potential
$V_{\rm PQ}$ in Eq.~(\ref{PQ-pot}) should be
shifted \cite{dllr1} by adding to it the
constant
\begin{eqnarray}
V_0&=&\frac{m_{3/2}^3 m_{\rm P}}{108\lambda}
\left(|A|+
\sqrt{|A|^2-12}\right)
\nonumber \\
& &\times\left[|A|\left(|A|+\sqrt{|A|^2-12}
\right)-24\right]
\label{eq:v0}
\end{eqnarray}
so that it vanishes at its absolute minima.

\section{THE PQ FIELD EVOLUTION}

\par
In the early universe, the PQ potential can
acquire sizable corrections from the SUSY
breaking caused by the presence of a finite
energy density \cite{lyth,crisis,drt95}. These
corrections are particularly important during
inflation and the subsequent inflaton
oscillations. We will ignore the $A$ term type
corrections \cite{drt95}. To
leading order, we then just obtain a correction
$\delta m_{\phi}^2=\gamma^2 H^2$ (assumed to be
positive) to the ${\rm mass}^2$ of the curvaton,
where $H$ is the Hubble parameter and the
dimensionless constant $\gamma$ can have
different values during inflation and inflaton
oscillations. We assume that
$\delta m_{\phi}^2$ is (somewhat) suppressed,
which can be \cite{noscale} the case for
specific (no-scale like) K\"{a}hler potentials.
For simplicity, during inflation where the
cancellation of $\delta m_{\phi}^2$ can, in
principle, be ``naturally'' arranged to be
exact (see fourth paper in
Ref.~\cite{noscale}), we take $\gamma=0$.
During inflaton oscillations, on the other
hand, we choose $\gamma=0.1$.

\par
After reheating, the universe is radiation
dominated and, thus, $H\simeq 1/2t\leq 1/
2t_{\rm reh}=\Gamma_{\rm infl}/2$, where $t$ is
the cosmic time and
$t_{\rm reh}=\Gamma_{\rm infl}^{-1}$ the time at
reheating with $\Gamma_{\rm infl}$ being the
inflaton decay width. The reheat temperature
$T_{\rm reh}$ is \cite{lectures}
\begin{equation}
T_{\rm reh}=\left(\frac{45}{2\pi^2 g_*}
\right)^{\frac{1}{4}}
(\Gamma_{\rm infl}m_{\rm P})^{\frac{1}{2}},
\label{eq:reheat}
\end{equation}
where $g_*$ is the effective number of massless
degrees of freedom ($g_*=228.75$ for the MSSM
spectrum). It should satisfy \cite{ekn} the
gravitino constraint: $T_{\rm reh}\leq 10^9~
{\rm GeV}$. The PQ potential can acquire
temperature corrections too both during the era
of inflaton oscillations (from the so-called
``new'' radiation \cite{reheat} emerging
from the decaying inflaton) and certainly after
reheating. It has been shown \cite{dllr1},
however, that these corrections are always
subdominant and can be ignored.

\par
The full effective scalar potential for the PQ
field in the early universe is given by
\begin{equation}
V_{\rm PQ}^{\rm eff}=V_{\rm PQ}+\frac{1}{2}
\gamma^2 H^2\phi^2+V_0,
\label{full-pot}
\end{equation}
whereas the full effective scalar potential $V$
relevant for our analysis is obtained
by adding to $V_{\rm PQ}^{\rm eff}$ the
potential for standard SUSY hybrid inflation
(see e.g. Ref.~\cite{lectures}). The evolution
of $\phi$ is generally governed by the classical
equation of motion
\begin{equation}
\ddot{\phi}+3H\dot{\phi}+V^{\prime}=0,
\label{field-eqn}
\end{equation}
where overdots and primes denote derivation
with respect to $t$ and the PQ field $\phi$
respectively. This equation holds
\cite{communication} during inflation for the
mean field in a region of fixed size bigger
than the de Sitter horizon provided that the
quantum fluctuations of $\phi$ do not
overshadow its classical kinetic energy. This
requires \cite{notari} that the value $\phi_f$
of $\phi$ at the end of inflation exceeds a
certain value $\phi_Q$ given by
\begin{equation}
V^{\prime}\sim\frac{3H_{\rm infl}^3}{2\pi},
\label{phiQ}
\end{equation}
where $H_{\rm infl}$ is the almost constant
Hubble parameter during inflation. We exclude
the ``quantum regime'' (i.e. $\phi_f<\phi_Q$),
where the calculation of the curvaton spectral
index is not \cite{iso} so clear.

\par
For $\phi$ to receive a super-horizon spectrum
of perturbations from inflation and thus be
able to act as curvaton, we must make sure that
this field is effectively massless, i.e.
$V^{\prime\prime}\leq H^2$, during (at least)
the last $50-60$ inflationary e-foldings. This
requirement, which also guarantees that $\phi$
is slowly rolling during the relevant part of
inflation and emerges with vanishing velocity
at the end of inflation, yields an upper bound
on $\phi_f$.

\par
The subsequent evolution of $\phi$ during
inflaton oscillations is given by
Eq.~(\ref{field-eqn}) with $H=2/3t$. One finds
\cite{dllr1} that, depending on the value of
$\phi_f$, the PQ system eventually enters into
damped oscillations about either the
trivial (local) minimum of $V_{\rm PQ}$ at
$\phi=0$ or one of its PQ (absolute) minima at
$\phi=\pm f_a$. Of course, values of $\phi_f$
leading to the trivial minimum must be excluded.

\par
The damped oscillations of $\phi$ continue even
after reheating, where $H$ becomes equal to
$1/2t$, until $\phi$ decays via the second
coupling in the RHS of Eq.~(\ref{W2}) into a
pair of Higgsinos provided that their mass
$\mu$ does not exceed half of the mass
$m_{\phi}$ of $\phi$ (see Ref.~\cite{dllr1}),
which is equal to
\begin{equation}
\frac{m_{3/2}}{\sqrt{3}}\left(|A|^2-
12\right)^{\frac{1}{4}}\left(|A|+\left(|A|^2-12
\right)^{\frac{1}{2}}\right)^{\frac{1}{2}}
\label{mphi}
\end{equation}
and is independent of $\lambda$. The decay time
of the PQ field is
$t_{\phi}=\Gamma_{\phi}^{-1}$, where
$\Gamma_{\phi}$ is its decay width, which has
been found \cite{dllr1} to be given by
\begin{equation}
\Gamma_{\phi}=\frac{\beta^2f_a^2}
{8\pi m_{\rm P}^2}m_{\phi}.
\label{Gphi}
\end{equation}
Note that the coherently oscillating PQ field
could evaporate \cite{evap} as a result of
scattering with particles in the thermal bath
before it decays into Higgsinos. However, one
can show \cite{dllr1} that, in our model, this
does not happen.

\section{THE CURVATURE PERTURBATION}

\par
The perturbation $\delta\phi$ acquired by
$\phi$ during inflation evolves at subsequent
times and, when $\phi$ settles into damped
quadratic oscillations about the PQ vacua,
yields a stable perturbation in the energy
density of this field. After the
PQ field decays, this perturbation is
transferred to radiation, which also carries a
curvature perturbation from the inflaton. So,
the total curvature perturbation is of mixed
origin.

\par
The total curvature perturbation on unperturbed
hypersurfaces is given by
\begin{equation}
\zeta=\frac{\delta\rho}{3(\rho+p)},
\end{equation}
where $\rho$ and $p$ are the total energy
density and pressure, and $\delta\rho$ the
total density perturbation. After the curvaton
decay, it becomes
\cite{curv3}
\begin{equation}
\zeta=(1-f)\zeta_i+f\zeta_c,
\label{zeta}
\end{equation}
where $\zeta_i=\delta\rho_r/4\rho_r$ and
$\zeta_c=\delta\rho_\phi/3\rho_\phi$ are the
partial curvature perturbations on spatial
hypersurfaces of constant curvature from the
inflaton and the curvaton respectively at the
curvaton decay with $\rho_r$ and $\rho_\phi$
being the radiation and $\phi$ energy densities
respectively, and $\delta\rho_r$ and $\delta
\rho_\phi$ the corresponding perturbations.
Also,
\begin{equation}
f=\frac{3\rho_{\phi}}{3\rho_{\phi}+4\rho_r}
\label{eq:f}
\end{equation}
evaluated at the time of the curvaton decay,
and $\zeta_c=2\delta\phi_0/3\phi_0$, where
$\phi_0$ is the amplitude of the quadratic
oscillations of $\phi$ and $\delta\phi_0$ the
perturbation in $\phi_0$ originating from the
perturbation $\delta\phi_f$ in $\phi_f$ at the
end of inflation.

\par
The comoving curvature perturbation
$\mathcal{R}_{\rm rad}$, for super-horizon
scales, is given by
\begin{equation}
(1-f)A_i\left(\frac{k}{H_0}\right)^{\nu_i}
\hat{a}_i+fA_c\left(\frac{k}{H_0}
\right)^{\nu_c}\hat{a}_c,
\label{R}
\end{equation}
where $k$ is the comoving (present physical)
wave number, $H_0=100h~\rm{km}~\rm{sec}^{-1}~
\rm{Mpc}^{-1}$ is the present Hubble parameter
and $\hat{a}_i$, $\hat{a}_c$ are
independent normalized Gaussian random
variables. Also, $A_i$ and $A_c$ are,
respectively, the amplitudes of $-\zeta_i$
and $-\zeta_c$ at $k=H_0$, and $\nu_i$ and
$\nu_c$ are the spectral tilts of the inflaton
and curvaton respectively. The corresponding
spectral indices are $n_i=1+2\nu_i$ and
$n_c=1+2\nu_c$. We do not consider running of
the spectral indices, since this is negligible
in our model.

\par
The amplitude $A_i$ of the partial curvature
perturbation from the inflaton is \cite{hier}
\begin{equation}
A_i=\left(\frac{2N_{Q}}{3}
\right)^{\frac{1}{2}}
\left(\frac{M}{m_{\rm P}}
\right)^2x_Q^{-1}y_Q^{-1}
\Lambda(x_Q^2)^{-1}
\label{ai}
\end{equation}
with
\begin{equation}
\Lambda(z)=(z+1)\ln(1+z^{-1})+(z-1)\ln(1-z^{-1}),
\label{lambda}
\end{equation}
\begin{equation}
y_Q^2=\int_{x_f^2}^{x_Q^2}\frac{dz}{z}
\Lambda(z)^{-1},~y_Q\geq 0~.
\label{yq}
\end{equation}
Here, $N_{Q}$ is the number of e-foldings
suffered by our present horizon during
inflation, $x_Q=S_Q/M$ with $S_Q$ being the
value of $\vert S\vert$ when our present
horizon crosses outside the inflationary
horizon, and $x_f=S_f/M$ with $S_f$ being the
value of $\vert S\vert$ at the end of
inflation.

\par
The slow-roll parameters for the inflaton are
\cite{lectures}
\begin{equation}
\epsilon_i=\left(\frac{\kappa^2m_{\rm P}}{8\pi^2M}
\right)^2z\Lambda(z)^2,
\label{epsiloni}
\end{equation}
\begin{eqnarray}
\eta_i&=&2\left(\frac{\kappa m_{\rm P}}{4\pi M}
\right)^2\Big[(3z+1)\ln(1+z^{-1})
\nonumber\\
& &+(3z-1)\ln(1-z^{-1})\Big],
\label{etai}
\end{eqnarray}
where $z=x^2$ with $x=\vert S\vert/M$. In the
presence of the curvaton, the slow-roll
conditions take the form $\epsilon,\vert\eta_i
\vert\leq 1$, where
\begin{equation}
\epsilon\equiv -\frac{\dot H}{H^2}=\epsilon_i+
\epsilon_c
\label{epsilon}
\end{equation}
with
\begin{equation}
\epsilon_c=\frac{1}{2}
m_{\rm P}^2\left(\frac{V^{\prime}}{V}\right)^2
\label{epsilonc}
\end{equation}
and $x_f$ corresponds to $\eta_i=-1$. For
$\kappa\ll 1$, $x_f$ is ``infinitesimally''
close to unity and we can put $x_f=1$ in
Eq.~(\ref{yq}). For larger values of $\kappa$,
inflation can terminate well before reaching
the instability point at $x=1$.

\par
Finally, $\kappa$, $N_Q$ are given by
\cite{lectures}
\begin{equation}
\kappa=\frac{2\pi}{\sqrt{N_Q}}
~y_Q~\frac{M}{m_{\rm P}},
\label{kappa}
\end{equation}
\begin{equation}
N_Q\simeq 55.9+\frac{2}{3}\ln\frac{
\kappa^{\frac{1}{2}}M}{10^{15}~{\rm GeV}}
+\frac{1}{3}\ln\frac{T_{\rm reh}}
{10^9~{\rm GeV}}
\label{NQ}
\end{equation}
and $n_i=1-6\epsilon+2\eta_i\simeq 1+2\eta_i$,
where $\epsilon$ and $\eta_i$ are evaluated at
the time $t_*$ when our present horizon scale
crosses outside the inflationary horizon.

\par
To calculate the amplitude $A_c$ of the partial
curvature perturbation from the curvaton, we
take the perturbation $\delta\phi_*=(H_*/2\pi)
\hat{a}_c$ acquired by $\phi$ from inflation at
$t_*$ ($H_*$ is the value of $H$ at $t_*$) and
follow its evolution until the end of inflation,
where we find $\delta\phi_f$. To this end, we
consider the equation of motion for $\phi$
during inflation (see Eq.~(\ref{field-eqn})) in
the slow-roll approximation, which holds for
the curvaton too:
\begin{equation}
3H\dot\phi+V^{\prime}=0.
\label{slow}
\end{equation}
Taking a small perturbation $\delta\phi$ of
$\phi$, this gives
\begin{equation}
3H\delta\dot\phi+3H^{\prime}\delta\phi\dot\phi+
V^{\prime\prime}\delta\phi=0,
\label{dphi1}
\end{equation}
which using Eq.~(\ref{slow}) and the Friedmann
equation, becomes
\begin{equation}
\delta\dot\phi+H(-\epsilon_c+\eta_c)\delta\phi=0,
\label{dphi2}
\end{equation}
where
\begin{equation}
\eta_c=m_{\rm P}^2\frac{V^{\prime\prime}}{V}.
\label{etac}
\end{equation}
Integration of Eq.~(\ref{dphi2}) from $t_*$
until the end of inflation (at time $t_f$)
yields
\begin{equation}
\delta\phi_f=\frac{H_*}{2\pi}\,\hat{a}_c
\exp{\int_0^{N_Q}(\epsilon_c-\eta_c)dN},
\label{dphif}
\end{equation}
where we used the relation $dN=-Hdt$ for the
number of e-foldings $N(k)=N_Q+\ln(H_0/k)$
suffered by the scale $k^{-1}$ during
inflation.

\par
For each $\phi_f$, we construct the perturbed
field $\phi_f+\delta\phi_f$. We then follow the
evolution of $\phi_f$ and $\phi_f+\delta\phi_f$
until the time $t_\phi$ of the curvaton decay
and evaluate $\delta\rho_\phi/\rho_\phi$ at
this time. The amplitude $A_c$ is given by
\begin{equation}
A_c\,\hat{a}_c=\frac{1}{3}\frac{\delta\rho_\phi}
{\rho_\phi}.
\label{Ac}
\end{equation}
We have found numerically that the perturbation
$\delta\phi_0$ in the amplitude of the
oscillating curvaton at $t_\phi$ is
proportional to $\delta\phi_f$. So $\zeta_c$
has the same spectral tilt as $\delta\phi_f$,
which can be found from Eq.~(\ref{dphif}):
\begin{equation}
\nu_c\equiv\frac{d\ln A_c}{d\ln k}=-\epsilon-
\epsilon_c+\eta_c,
\label{nuc}
\end{equation}
where we used the relation $d\ln k=Hdt$, and
$\epsilon$, $\epsilon_c$ and $\eta_c$ are
evaluated at $t_*$. The curvaton spectral
index is $n_c=1-2\epsilon-2\epsilon_c+
2\eta_c\simeq 1+2\eta_c$.

\section{ISOCURVATURE PERTURBATIONS}

\par
At reheating, gravitinos are thermally
produced and decay well after the big bang
nucleosynthesis (BBN) into a particle and a
sparticle, which subsequently turns into a
(stable) lightest sparticle (LSP) contributing
to the relic abundance of cold dark matter
(CDM) in the universe. For simplicity, we
assume that there are no
thermally produced LSPs. Baryons are produced
via a primordial leptogenesis \cite{lepto}
occurring \cite{leptoinf} at reheating. So,
both the LSPs and the baryons originate from
reheating and, thus, inherit the partial
curvature perturbation of the inflaton, i.e.
\begin{equation}
\zeta_{\rm LSP}=\zeta_B=\zeta_i.
\label{zetaLSP}
\end{equation}
The isocurvature perturbation of the LSPs and
the baryons is then
\begin{equation}
\mathcal{S}_{{\rm LSP}+B}\equiv 3(
\zeta_{{\rm LSP}+B}-\zeta)=3f(\zeta_i-\zeta_c),
\label{SLSP1}
\end{equation}
where $\zeta_{{\rm LSP}+B}=\zeta_{\rm LSP}=
\zeta_B$ is the partial curvature perturbation
of the LSPs and the baryons and we used
Eq.~(\ref{zeta}). Here, we assume that the
curvature perturbation in radiation
($\zeta_\gamma$) practically coincides with the
total curvature perturbation. This corresponds
to a negligible neutrino isocurvature
perturbation, which is \cite{curv3} the case
provided that leptogenesis takes place well
before the curvaton decays or dominates the
energy density. Applying the definitions which
follow Eq.~(\ref{R}),
$\mathcal{S}_{{\rm LSP}+B}$ becomes equal to
\begin{equation}
-3fA_i\left(\frac{k}{H_0}\right)^{\nu_i}
\hat{a}_i+3fA_c\left(\frac{k}{H_0}
\right)^{\nu_c}\hat{a}_c.
\label{SLSP2}
\end{equation}

\par
The model contains axions which can also
contribute to CDM. They are produced at the QCD
phase transition well after the curvaton decay
and carry an isocurvature perturbation
\begin{equation}
\mathcal{S}_a=A_a\left(\frac{k}{H_0}
\right)^{\nu_a}\hat{a}_a,
\label{Sa}
\end{equation}
where $A_a$ is its amplitude at the present
horizon scale, $\nu_a$ its spectral tilt
(giving the spectral index $n_a=1+2\nu_a$) and
$\hat{a}_a$ a normalized Gaussian random
variable independent from $\hat{a}_i$
and $\hat{a}_c$.

\par
The amplitude $A_a$ is given by \cite{dllr1}
\begin{equation}
A_a=\frac{H_*}{\pi\vert\theta\vert\phi_*},
\label{Aa}
\end{equation}
where $\theta$ is the initial misalignment
angle, i.e. the phase of the complex PQ field
during inflation, and $\phi_*$ is the value of
$\phi$ at $t_*$. In our case, $\theta$ lies
\cite{dllr1} in the interval $[-\pi/6,\pi/6]$
and is determined from the total CDM abundance
$\Omega_{\rm CDM}h^2$ which is the sum of the
relic abundance \cite{km}
\begin{equation}
\Omega_{\rm LSP}h^2\simeq 0.0074\left(\frac
{m_{\rm LSP}}{200~{\rm GeV}}\right)\left(
\frac{T_{\rm reh}}{10^{9}~{\rm GeV}}\right)
\label{OmegaLSP}
\end{equation}
of the LSPs and the relic axion abundance
\cite{axion}
\begin{equation}
\Omega_ah^2\simeq\theta^2\left(\frac{f_a}
{10^{12}~{\rm GeV}}\right)^{1.175}.
\label{Omegaa}
\end{equation}
Here $\Omega_j=\rho_j/\rho_c$ with $\rho_j$
being the present energy density of the $j$th
species and $\rho_c$ the present critical
energy density of the universe, and
$m_{\rm LSP}$ is the LSP mass. The spectral
tilt $\nu_a$ is evaluated by observing
\cite{iso} that $A_a$ depends on the scale only
through $H_*/\phi_*$. We find
\begin{equation}
\nu_a=-\epsilon+\frac{m_{\rm P}}{\phi_*}
\frac{V^\prime}{V}m_{\rm P}=
-\epsilon+\frac{m_{\rm P}}{\phi_*}
\sqrt{2\epsilon_c},
\label{nua}
\end{equation}
where the $\epsilon$ and $\epsilon_c$ are
evaluated at $t_*$. The axion spectral index is
$n_a\simeq 1+2m_{\rm P}\sqrt{2\epsilon_c}/
\phi_*$.

\par
Combining Eqs.~(\ref{SLSP2}) and (\ref{Sa}), we
find that the total isocurvature perturbation is
\cite{dllr1}
\begin{equation}
\mathcal{S}_{\rm rad}=\frac{\Omega_{{\rm LSP}+B}}
{\Omega_m}\mathcal{S}_{{\rm LSP}+B}+\frac{\Omega_a}
{\Omega_m}\mathcal{S}_a,
\label{total}
\end{equation}
where we used the definitions
$\Omega_{{\rm LSP}+B}\equiv\Omega_{\rm LSP}+
\Omega_B$ and $\Omega_m\equiv\rho_m/\rho_c=
\Omega_{{\rm LSP}+B}+\Omega_a$ with $\rho_m$
being the total matter density at present.

\section{THE CMBR POWER SPECTRUM}

\par
We will now calculate the total CMBR angular
power spectrum $C_\ell$ at the customarily used
\cite{pivot} pivot scale $k_{\rm P}=0.05~
{\rm Mpc^{-1}}$. We thus define the amplitudes
of the partial curvature perturbations from the
inflaton ($A_{{\rm P},i}$) and the curvaton
($A_{{\rm P},c}$), and the amplitude of the
isocurvature perturbation in the axions
($A_{{\rm P},a}$) at $k=k_{\rm P}$:
\begin{eqnarray*}
A_{{\rm P},i}=A_i\left(\frac{k_{\rm P}}{H_0}
\right)^{\nu_i},~
A_{{\rm P},c}=A_c\left(\frac{k_{\rm P}}{H_0}
\right)^{\nu_c},
\end{eqnarray*}
\begin{equation}
A_{{\rm P},a}=A_a\left(\frac{k_{\rm P}}{H_0}
\right)^{\nu_a}.
\label{AP}
\end{equation}

\par
The amplitude squared of the adiabatic
perturbation at $k_{\rm P}$ is then given by
\begin{equation}
R^2=\langle\mathcal{R}_{\rm rad}
\mathcal{R}_{\rm rad}\rangle=R_i^2+R_c^2,
\label{R2}
\end{equation}
where $\mathcal{R}_{\rm rad}$ is evaluated at
$k_{\rm P}$, and the inflaton ($R_i^2$) and
curvaton ($R_c^2$) contributions to $R^2$ are
\begin{equation}
R_i^2=(1-f)^2A_{{\rm P},i}^2~~{\rm and}~~R_c^2=
f^2A_{{\rm P},c}^2.
\label{Ric}
\end{equation}
The curvaton fractional contribution to $R$ is
\begin{equation}
F^{\rm ad}_c=\frac{R_c}{R}.
\label{Fc}
\end{equation}
The amplitude squared of the isocurvature
perturbation at $k_{\rm P}$ is found from
Eq.~(\ref{total}) to be
\begin{equation}
S^2=\langle\mathcal{S}_{\rm rad}
\mathcal{S}_{\rm rad}\rangle=S_i^2+S_c^2+S_a^2,
\label{S2}
\end{equation}
where
\begin{eqnarray*}
S_i^2=9f^2\left(\frac{\Omega_{{\rm LSP}+B}}
{\Omega_m}\right)^2A_{{\rm P},i}^2,
\end{eqnarray*}
\begin{eqnarray*}
S_c^2=9f^2\left(\frac{\Omega_{{\rm LSP}+B}}
{\Omega_m}\right)^2A_{{\rm P},c}^2,
\end{eqnarray*}
\begin{equation}
S_a^2=\left(\frac{\Omega_a}{\Omega_m}\right)^2
A_{{\rm P},a}^2
\label{Sica}
\end{equation}
are, respectively, the inflaton, curvaton,
and axion contributions to this amplitude
squared. The cross correlation between the
adiabatic and isocurvature perturbation at
$k_{\rm P}$ is
\begin{equation}
C=\langle\mathcal{R}_{\rm rad}
\mathcal{S}_{\rm rad}\rangle=C_i+C_c,
\label{C}
\end{equation}
where
\begin{equation}
C_i=-R_iS_i~~{\rm and}~~C_c=R_cS_c
\label{Cic}
\end{equation}
are the contributions from the inflaton and the
curvaton respectively. The axions do not
contribute to the cross correlation (and the
amplitude of the adiabatic perturbation). Also,
the isocurvature perturbation from the inflaton
is fully anti-correlated with the corresponding
adiabatic perturbation, whereas the
isocurvature perturbation from the curvaton is
fully correlated with the adiabatic
perturbation from it. So the overall correlation
is mixed. It is thus useful to define \cite{iso3}
the dimensionless cross correlation $\cos\Delta$
and the entropy-to-adiabatic ratio $B$ at
$k_{\rm P}$:
\begin{equation}
\cos\Delta=\frac{C}{RS}~~{\rm and}~~B=\frac{S}{R}.
\label{cosDB}
\end{equation}

\par
The total CMBR power spectrum is given, in the
notation of Ref.~\cite{thesis}, by the
superposition
\begin{equation}
C_\ell=C_\ell^{\rm ad}+C_\ell^{\rm is}+
C_\ell^{\rm cc},
\label{Cell}
\end{equation}
where
\begin{equation}
C_\ell^{\rm ad}=R_i^2C_\ell^{{\rm ad},n_i}+
R_c^2C_\ell^{{\rm ad},n_c},
\label{Cellad}
\end{equation}
\begin{equation}
C_\ell^{\rm is}=S_i^2C_\ell^{{\rm is},n_i}+
S_c^2C_\ell^{{\rm is},n_c}+
S_a^2C_\ell^{{\rm is},n_a},
\label{Cellis}
\end{equation}
\begin{equation}
C_\ell^{\rm cc}=C_iC_\ell^{{\rm cc},n_i}+
C_cC_\ell^{{\rm cc},n_c}.
\label{Cellcc}
\end{equation}
These relations hold for the temperature (TT),
E-polarization and temperature-polarization
(TE) cross correlation angular power spectra.

\par
As an indicative (approximate) criterion to get
a first rough feeling on the possible
compatibility of our model with the data, we
apply the cosmic microwave background explorer
(COBE) constraint \cite{cobe10} on the TT power
spectrum $C^{\rm TT}_\ell$:
\begin{equation}
\ell(\ell+1)C^{\rm TT}_\ell/2\pi\big
\vert_{\ell=10}\approx 1.05\times 10^{-10}
\label{ell10}
\end{equation}
with
\begin{eqnarray}
C^{\rm TT}_\ell&=&\frac{2\pi^2}{25}
\big[(R_i^2+4S_i^2-4C_i)f(n_i,\ell)
\nonumber\\
& &+(R_c^2+4S_c^2-4C_c)f(n_c,\ell)
\nonumber\\
& &+4S_a^2f(n_a,\ell)\big],
\label{Cobe}
\end{eqnarray}
which is accurate to about $10-20\%$ in the
Sachs-Wolfe (SW) plateau, i.e. for $\ell\leq
20$ (see e.g. Ref.~\cite{thesis}). Here the
function $f(n,\ell)$ is equal to
\begin{equation}
(\eta_0k_{\rm P})^{1-n}
\frac{\Gamma(3-n)
\Gamma(\ell-\frac{1}{2}+\frac{n}{2})}{2^{3-n}
\Gamma^2(2-\frac{n}{2})\Gamma(\ell+\frac{5}{2}
-\frac{n}{2})}
\label{fnell}
\end{equation}
with $\eta_0=2H_0^{-1}\simeq 8.33\times 10^3~
{\rm Mpc}$ being the present conformal time.

\section{NUMERICAL CALCULATIONS}

\subsection{The evolution of the curvaton}

\par
We are now ready to proceed to the numerical
study of the evolution of the PQ field during
and after inflation. To this end, we fix
$\kappa$ in the superpotential $\delta W_1$ in
Eq.~(\ref{W1}). Then, for any given value of
$A_i$, we solve Eqs.~(\ref{ai}), (\ref{kappa})
and (\ref{NQ}), where $x_f$ is the solution of
$\eta_i=-1$ with $\eta_i$ given by
Eq.~(\ref{etai}) and $T_{\rm reh}$, which
enters Eq.~(\ref{NQ}), is taken equal to
$10^9~{\rm GeV}$ by saturating the gravitino
bound \cite{ekn}. We thus determine the mass
parameter $M$, which is the $G_{LR}$-breaking
VEV. Subsequently, we find the (almost
constant) inflationary Hubble parameter
$H_{\rm infl}=\kappa M^2/\sqrt{3}m_{\rm P}$.

\par
For any given $A_i$, we take a value $\phi_f$
of $\phi$ at the end of inflation (at $t_f=2/
3H_{\rm infl}$). We then solve the slow-roll
classical equation of motion for the field
$\phi$ during
inflation going backwards in time ($t\leq t_f$)
by taking $m_{3/2}=300~{\rm GeV}$, $\vert A
\vert=5$ and a fixed $\lambda$ ($\sim 10^{-4}$)
in the PQ potential. Finally, we put $\gamma=0$
during inflation. We find that, as we move
backwards in time, $\phi$ increases and becomes
infinite at a certain moment. The number of
e-foldings elapsed from this moment until the end
of inflation is finite providing an upper bound
$N_{\rm max}$ on the number of e-foldings which
is compatible with $\phi=\phi_f$ at $t_f$.

\par
To understand this behavior, we approximate the
potential $V$ by $V\simeq\lambda^2\phi^6/32
m_{\rm P}^2$, which holds for $\phi\rightarrow
\infty$. The slow-roll equation for $\phi$
during inflation can then be solved
analytically yielding
\begin{equation}
\phi\simeq\frac{\phi_f}
{\left(1+\frac{\lambda^2\phi_f^4}
{4m_{\rm P}^2H_{\rm infl}}\Delta t
\right)^{\frac{1}{4}}},
\label{analytic}
\end{equation}
where $\Delta t=t-t_f\leq 0$. As $\Delta t
\rightarrow\Delta t_{\rm min}\equiv
-4m_{\rm P}^2H_{\rm infl}/\lambda^2\phi_f^4$,
$\phi\rightarrow\infty$, which implies that the
maximal number of e-foldings allowed for a
given $\phi_f$ is $N_{\rm max}\simeq
-H_{\rm infl}\Delta t_{\rm min}=4m_{\rm P}^2
H_{\rm infl}^2/\lambda^2\phi_f^4$.

\begin{figure}[tb]
\centering
\includegraphics[width=0.83\linewidth]
{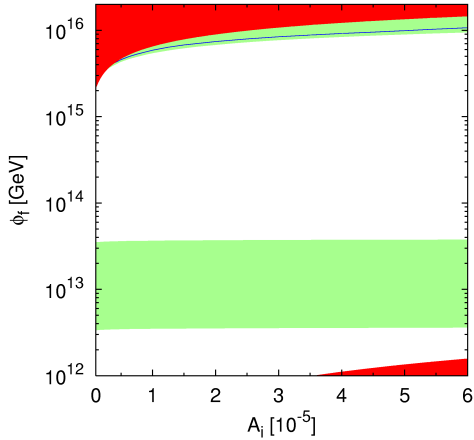}
\caption{The two green/lightly shaded bands
in the $A_i-\phi_f$ plane which lead to the
PQ vacua at $t_{\phi}$ for $\kappa=3\times
10^{-3}$, $\lambda=10^{-4}$ (model A). The
white (not shaded) areas lead to the trivial
vacuum and are thus excluded. The upper
red/dark shaded area is excluded by the
requirement that, at $t_*$,
$V^{\prime\prime}\leq H_{\rm infl}^2$, while
the lower one corresponds to the quantum
regime. The blue/solid line shows the values
of $A_i$, $\phi_f$ which approximately
reproduce the correct value of the CMBR large
scale temperature anisotropy, as measured by
COBE.}
\label{fig:model-4}
\end{figure}

\begin{figure}[tb]
\centering
\includegraphics[width=0.83\linewidth]
{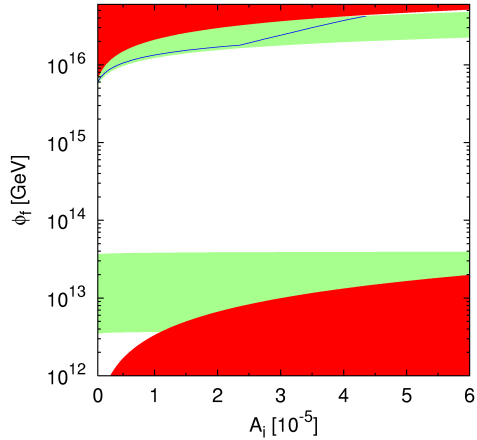}
\caption{The two green/lightly shaded bands
in the $A_i-\phi_f$ plane which lead to the
PQ vacua at $t_{\phi}$ for $\kappa=3\times
10^{-2}$, $\lambda=10^{-4}$ (model B). The
notation is the same as in
Fig.~\ref{fig:model-4}.}
\label{fig:model-5}
\end{figure}

\par
We must first require that $N_{\rm max}\geq
N_Q$. The time $t_*$ is then found from
$N_Q=H_{\rm infl} (t_f-t_*)$. Furthermore, we
demand that, at $t_*$, $V^{\prime\prime}\leq
H_{\rm infl}^2$, which ensures that this
inequality holds for all times between $t_*$
and $t_f$. This guarantees that $\phi$ is
effectively massless during inflation and can
act as curvaton. It also justifies the use
of the slow-roll approximation and ensures that
the velocity of $\phi$ at the end of inflation
is negligible. This masslessness requirement
yields an upper bound on $\phi_f$ for every
$A_i$ and fixed $\kappa$ and $\lambda$.
The excluded region in the $A_i-\phi_f$ plane
for fixed $\kappa$ and $\lambda$ is depicted
as a red/dark shaded area in the upper part of
this plane. In Figs.~\ref{fig:model-4} and
\ref{fig:model-5}, we show this upper red/dark
shaded area for $\kappa=3\times 10^{-3}$,
$\lambda=10^{-4}$ (model A) and $\kappa=3\times
10^{-2}$, $\lambda=10^{-4}$ (model B)
respectively. The lower red/dark shaded area
corresponds to the quantum regime and is also
excluded.

\par
We start from any given $\phi_f$ at $t_f$ not
excluded by the above considerations and
assume zero time derivative of $\phi$ at $t_f$.
We then follow the evolution of $\phi$ for
$t\geq t_f$ by solving the equation
of motion in Eq.~(\ref{field-eqn}) with
$H=2/3t$ for $t_f\leq t\leq t_{\rm reh}$ and
$H=1/2(t-t_{\rm reh}/4)$ for $t_{\rm reh}
\leq t\leq t_\phi$. The time of reheating
$t_{\rm reh}$ is found from
Eq.~(\ref{eq:reheat}) with $g_*=228.75$ and
the curvaton decay time $t_\phi$ from
Eq.~(\ref{Gphi}) with $\beta$ from
Eq.~(\ref{mu}), where we put $\mu=300~
{\rm GeV}$. We take $\gamma=0.1$ after the end
of inflation. We find that, for fixed $\kappa$
and $\lambda$, there exist two bands in the
$A_i-\phi_f$ plane leading to the PQ vacua at
$t_\phi$. They are depicted as an upper and a
lower green/lightly shaded band (see
Figs.~\ref{fig:model-4} and \ref{fig:model-5}).
The white (not shaded) areas in the
$A_i-\phi_f$ plane lead to the false (trivial)
minimum at $\phi=0$ and thus must be excluded.

\subsection{The calculation of $C_\ell$}

\par
For any fixed $\kappa$ and $\lambda$, we take a
grid of values of $A_i$ and $\phi_f$ which span
the corresponding upper or lower green/lightly
shaded band. For each point on this grid, we
consider $\phi_f$ and its perturbed value
$\phi_f+\delta\phi_f$ and follow the evolution
of both these fields until curvaton decay (at
$t_\phi$), where we evaluate the amplitude of
$\delta\rho_\phi/\rho_\phi$. The amplitude
$A_c$ is then given by Eq.~(\ref{Ac}). For the
Monte Carlo (MC) analysis, we use $A_i$ and
$A_c$ (rather than $A_i$ and $\phi_f$) as base
parameters and limit our grid to $A_i\leq 6
\times 10^{-5}$. The indices $n_i$ and $n_c$
for each point on the grid are found from
Eqs.~(\ref{epsiloni}) and (\ref{etai}) applied
at $x=x_Q$, and Eqs.~(\ref{epsilonc}),
(\ref{etac}) and (\ref{nuc}) applied at $t_*$.
The fraction $f$ in Eq.~(\ref{eq:f}) is
evaluated at $t_\phi$. The amplitude $A_a$ of
the isocurvature perturbation in the axions is
calculated from Eq.~(\ref{Aa}) with the initial
misalignment angle $\theta$ evaluated, for any
given total CDM abundance $\Omega_{\rm CDM}h^2
=\Omega_{\rm LSP}h^2+\Omega_ah^2$, by using
Eqs.~(\ref{OmegaLSP}) and (\ref{Omegaa}) with
$m_{\rm LSP}=200~{\rm GeV}$. For our choice of
parameters, the LSP relic abundance is fixed
($\simeq 0.0074$). The spectral index for
axions is found from Eq.~(\ref{nua}).

\par
In summary, for any fixed $\kappa$ and
$\lambda$, we take a grid in the variables
$A_i$ and $A_c$ covering the upper or lower
green/lightly shaded band. For any
$\Omega_Bh^2$ and $\Omega_ah^2$, we then
calculate the amplitudes squared of the
adiabatic and isocurvature perturbations from
Eqs.~(\ref{R2}), (\ref{Ric}) and
Eqs.~(\ref{S2}), (\ref{Sica}) respectively,
the cross correlation amplitude from
Eqs.~(\ref{C}), (\ref{Cic}) and the total CMBR
TT and TE power spectra via
Eqs.~(\ref{Cell})$-$(\ref{Cellcc}). The
curvaton fractional contribution to the
adiabatic amplitude $F^{\rm ad}_c$, the
dimensionless cross correlation $\cos\Delta$
and the entropy-to-adiabatic ratio $B$ are
found from Eqs.~(\ref{Fc}) and (\ref{cosDB}).
In computing $M$, $A_c$ and $A_a$, we fix
the present Hubble parameter to the Hubble
space telescope (HST) central value $H_0=72~
\rm{km}~\rm{sec}^{-1}~\rm{Mpc}^{-1}$
\cite{hst}, which has an impact less than 3\%
on our results. Clearly, we do allow $H_0$ to
vary in the MC analysis.

\par
Before deploying the full MC machinery to
derive constraints on the parameters, it is
instructive to obtain a first rough idea about
the viability of our model by using the
approximate expressions in Eqs.~(\ref{Cobe})
and (\ref{fnell}) for the temperature SW
plateau and requiring that the COBE constraint
in Eq.~(\ref{ell10}) is fulfilled. To this end,
we take $\Omega_Bh^2=0.0224$ and
$\Omega_{\rm CDM}h^2=0.1126$, which are the
best-fit values from WMAP \cite{wmap1}.  We
find that the COBE constraint cannot be
satisfied in the lower green/lightly shaded
band in the $A_i-\phi_f$ plane for any
$\kappa$ and $\lambda$. The reason is that the
relatively low values of $\phi_*$ in this band
combined with the sizable relic abundance of
the axions leads to an unacceptably large axion
contribution to the RHS of Eq.~(\ref{Cobe}). In
the upper green/lightly shaded band in the
$A_i-\phi_f$ plane, on the contrary, the COBE
constraint is easily satisfied. The resulting
solution is depicted by a blue/solid line (see
Figs.~\ref{fig:model-4} and \ref{fig:model-5}).

\subsection{The MC analysis}

\par
We proceed to the full MC analysis confronting
the predictions of our model with the CMBR
data. We use a version of the Markov chain MC
package \textsc{cosmomc} \cite{cosmomc} as
described in Ref.~\cite{lewis}. The adiabatic
and isocurvature CMBR transfer functions are
computed in two successive calls similarly to
Ref.~\cite{Valiviita}. For fixed $\kappa$ and
$\lambda$, the initial conditions are fully
specified by $A_i$ and $A_c$. The MC sampling
takes as free parameters the $A_i$ and $A_c$,
the present baryon and axion abundances
$\omega_B\equiv\Omega_Bh^2$ and
$\omega_a\equiv\Omega_ah^2$, the present
dimensionless Hubble parameter $h$, and the
redshift $z_r$ at which the reionization
fraction is a half. All other derived
quantities are computed from the above
parameter set. In particular, the total CDM
abundance is $\omega_{\rm CDM}\equiv
\Omega_{\rm CDM}h^2=\omega_{\rm LSP}+\omega_a$,
where $\omega_{\rm LSP}\equiv\Omega_{\rm LSP}
h^2\approx 0.0074={\rm constant}$. Also, since
we take flat cosmologies only, the
cosmological constant energy density
$\Omega_\Lambda$ (in units of $\rho_c$) is
a derived parameter, i.e. $\Omega_\Lambda=1-
(\omega_{\rm CDM}+\omega_B)/h^2$. The
gravitational waves are negligible.
In summary, for fixed $\kappa$ and $\lambda$,
our parameter space is six dimensional:
\begin{equation}
\{\omega_B,\omega_a,h,z_r,A_i,A_c\}.
\end{equation}

\begin{figure}[tb]
\centering
\includegraphics[width=\linewidth]
{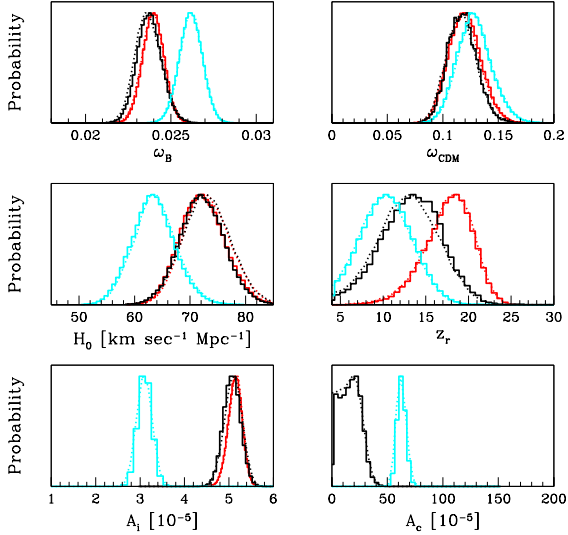}
\caption{1D marginalized
posterior distribution for model A (black solid
line) and model B (cyan/light gray solid line)
with only the upper green/lightly shaded bands
included. The red/medium gray line is for the
pure inflaton case with $n_i=1$, plotted here
for comparison. Model A is quite close to the
pure inflaton case. Model B displays a
preference for non-zero curvaton contribution.
Its quality of fit is, however, poorer. We also
display as dotted smoothed curves the values of
the mean posterior. All the curves are
normalized at their peak value.}
\label{fig:1dbase}
\end{figure}

\begin{figure}[tb]
\centering
\includegraphics[width=\linewidth]
{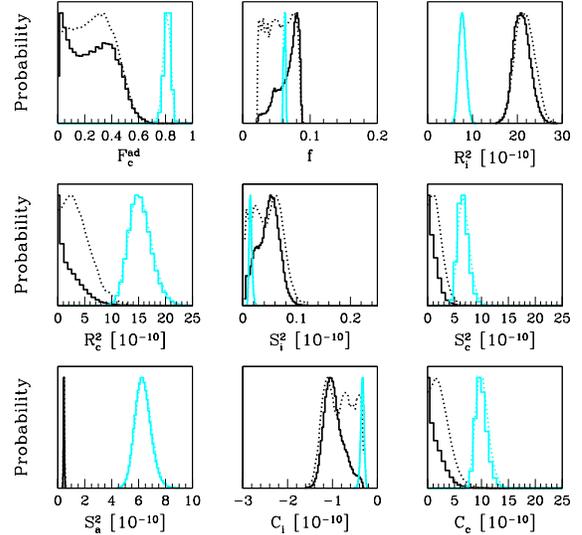}
\caption{As in Fig.~\ref{fig:1dbase}, but for
(most of) the derived parameters (the pure
inflaton model is not included). The
adiabatic amplitude in model B is dominated by
the curvaton ($F_c^{\rm ad}\approx 0.8$).}
\label{fig:1dderiv}
\end{figure}

\par
We compare the predicted CMBR TT and TE power
spectra to the WMAP first-year data
\cite{wmap1} with the routine for computing
the likelihood supplied by the WMAP team
\cite{Verde}. At a smaller angular scale, we
add the cosmic background imager (CBI)
\cite{cbi} and the decorrelated arcminute
cosmology bolometer array receiver (ACBAR)
\cite{acbar} band powers as well. We use flat
top-hat priors on the base parameters
\begin{equation}
\omega_B, \omega_a, z_r, A_i, A_c.
\end{equation}
The limits of the top-hat priors do not matter
for parameter estimation, as long as the
posterior density is negligible near the
limits. However, the prior range of the
accessible parameter space plays an important
role in computing the Bayes factor for model
testing. We limit the maximum range of $h$ by
imposing a top-hat prior $0.40<h<1.00$ and we
use the HST result \cite{hst}
\begin{equation}
\like^{\rm HST}\propto\exp\left(\frac{h-h_0}
{2\sigma}\right)^2,
\end{equation}
where $h_0=0.72$ and $\sigma=0.08$, by
multiplying the likelihood function for the
CMBR data by the above Gaussian likelihood.
By trying different priors, we find \cite{iso}
that the broad lines of the constraints for the
PQ model are robust with respect to the choice
of non-informative priors.

\par
We will examine in some detail the models A and
B, which are representative cases of the two
possible behavior patterns of our PQ model. We
first consider the upper green/lightly shaded
band of these models. The 1-dimensional (1D)
marginalized posterior distributions for the
base and most of the derived parameters are
plotted, respectively, in
Figs.~\ref{fig:1dbase} and \ref{fig:1dderiv}.

\par
In the upper band of model A, the power spectra
are dominated by the adiabatic inflaton
contribution. The quality of the fit, as
expressed by the maximum likelihood, is
slightly better than for the pure inflaton case
with $n_i=1$, which is not surprising since the
curvaton contribution plays a modest role. The
CMBR data yield the bound $A_c< 43.2\times
10^{-5}$ at $95\%$ confidence level. The total
TT power on large scales is slightly larger than
the pure adiabatic part, which increases the
height of the SW plateau relative to the height
of the first adiabatic peak. This mimics the
impact of a larger $z_r$, and explains why
model A prefers a later reionization than the
pure inflaton case.

\par
The upper band of model B exhibits a
preference for a non-vanishing curvaton
amplitude with very high significance
($A_c=62.8\pm 10.4$). The overall quality of
the fit is though worse than for model A (upper
band) because this model does not reproduce
with enough precision the shape of the first
acoustic peak. Furthermore, this model shows a
preference for a rather high baryon abundance
($\omega_B\approx 0.026$), which is in strong
tension with the value indicated by BBN
together with observations of the light
elements abundance, which typically yields
$\omega_B\approx 0.020\pm 0.002$ \cite{Burles}.

\par
For the lower band of model A, the quality of
the best fit is poor, whereas the lower band of
model B is incapable of producing a spectrum in
reasonable agreement with the data. Thus, the
lower bands of these models are excluded.

\par
So far we have derived parameter constraints
for models A and B from the data. The next
question is to compare the upper bands of these
two models with the pure inflaton
scale-invariant model and decide which of these
three models is most favored by data. Model
comparison is a different issue than parameter
extraction and it can very well be that it
arrives at a different conclusion. Indeed, it
can be that the estimated value of a parameter
under a model $\mdl_1$ is far from the null
value predicted by model $\mdl_2$, but $\mdl_1$
is disfavored against $\mdl_2$ by Bayesian
model testing. This is the case for $A_c$ in
model B (upper band) compared to the pure
inflaton model.

\par
To apply Bayesian model testing, we compute the
Bayes factor $B$ for models A and B (upper
bands) comparing each of them to the pure
inflaton model with $n_i=1$. We find \cite{iso}
that $\ln B =-3.9$ and $-17.1$ in the two cases
respectively. Thus, model A (upper band) is
disfavored against the pure inflaton model with
odds of $50$ to $1$, while model B (upper band)
with odds of $10^7$ to $1$. So, model B (upper
band), which predicts a non-zero curvaton
contribution to the adiabatic perturbation, is
strongly disfavored. On the contrary, model A
(upper band), which does not require such a
contribution, is only mildly disfavored.
However, in view of the fact that it addresses
many issues lying outside the scope of the pure
inflaton model, such as the strong CP and $\mu$
problems, the generation of the BAU and the
nature of CDM, we can by no means reject it.

\section{CONCLUSIONS}

\par
We studied a SUSY GUT model solving the strong
CP and $\mu$ problems via a PQ symmetry and
leading to standard hybrid inflation. The PQ
field acts as curvaton contributing to the
curvature perturbation together with the
inflaton. The model predicts isocurvature
perturbation too of mixed correlation with the
adiabatic one. We confronted the predictions of
the model with the CMBR data from WMAP, CBI and
ACBAR and found that a non-zero curvaton
contribution to the adiabatic perturbation is
not favored.

\section*{ACKNOWLEDGEMENT}
\par
This work was supported by European Union
under the contract MRTN-CT-2004-503369.

\def\ijmp#1#2#3{{Int. Jour. Mod. Phys.}
#1~(#2)~#3}
\def\plb#1#2#3{{Phys. Lett. B }#1~(#2)~#3}
\def\zpc#1#2#3{{Z. Phys. C }#1~(#2)~#3}
\def\prl#1#2#3{{Phys. Rev. Lett.}
#1~(#2)~#3}
\def\rmp#1#2#3{{Rev. Mod. Phys.}
#1~(#2)~#3}
\def\prep#1#2#3{{Phys. Rep. }#1~(#2)~#3}
\def\prd#1#2#3{{Phys. Rev. D }#1~(#2)~#3}
\def\npb#1#2#3{{Nucl. Phys. B }#1~(#2)~#3}
\def\npps#1#2#3{{Nucl. Phys. B (Proc. Suppl.)}
#1~(#2)~#3}
\def\mpl#1#2#3{{Mod. Phys. Lett.}
#1~(#2)~#3}
\def\arnps#1#2#3{{Annu. Rev. Nucl. Part. Sci.}
#1~(#2)~#3}
\def\sjnp#1#2#3{{Sov. J. Nucl. Phys.}
#1~(#2)~#3}
\def\jetp#1#2#3{{JETP Lett. }#1~(#2)~#3}
\def\app#1#2#3{{Acta Phys. Polon.}
#1~(#2)~#3}
\def\rnc#1#2#3{{Riv. Nuovo Cim.}
#1~(#2)~#3}
\def\ap#1#2#3{{Ann. Phys. }#1~(#2)~#3}
\def\ptp#1#2#3{{Prog. Theor. Phys.}
#1~(#2)~#3}
\def\apjl#1#2#3{{Astrophys. J. Lett.}
#1~(#2)~#3}
\def\n#1#2#3{{Nature }#1~(#2)~#3}
\def\apj#1#2#3{{Astrophys. J.}
#1~(#2)~#3}
\def\anj#1#2#3{{Astron. J. }#1~(#2)~#3}
\def\apjs#1#2#3{{Astrophys. J. Suppl.}
#1~(#2)~#3}
\def\mnras#1#2#3{{Mon. Not. Roy. Astron. Soc.}
#1~(#2)~#3}
\def\grg#1#2#3{{Gen. Rel. Grav.}
#1~(#2)~#3}
\def\s#1#2#3{{Science }#1~(#2)~#3}
\def\baas#1#2#3{{Bull. Am. Astron. Soc.}
#1~(#2)~#3}
\def\ibid#1#2#3{{ibid. }#1~(#2)~#3}
\def\cpc#1#2#3{{Comput. Phys. Commun.}
#1~(#2)~#3}
\def\astp#1#2#3{{Astropart. Phys.}
#1~(#2)~#3}
\def\epjc#1#2#3{{Eur. Phys. J. C}
#1~(#2)~#3}
\def\nima#1#2#3{{Nucl. Instrum. Meth. A}
#1~(#2)~#3}
\def\jhep#1#2#3{{J. High Energy Phys.}
#1~(#2)~#3}
\def\lnp#1#2#3{{Lect. Notes Phys.}
#1~(#2)~#3}
\def\appb#1#2#3{{Acta Phys. Polon. B}
#1~(#2)~#3}
\def\fcp#1#2#3{{Fund. Cos. Phys.}
#1~(#2)~#3}
\def\ptps#1#2#3{{Prog. Theor. Phys. Suppl.}
#1~(#2)~#3}
\def\ss#1#2#3{{Statist. Sci.}
#1~(#2)~#3}
\def\jasa#1#2#3{{J. Amer. Statist. Assoc.}
#1~(#2)~#3}
\def\jcap#1#2#3{{J. Cosmol. Astropart. Phys.}
#1~(#2)~#3}
\def\bm#1#2#3{{Biometrika}
#1~(#2)~#3}
\def\ams#1#2#3{{Ann. Math. Stat.}
#1~(#2)~#3}


\begin{thebibliography}{9}

\bibitem{llbook}
A.R. Liddle and D.H. Lyth, Cosmological
inflation and large-scale structure, Cambridge
Univ. Press, Cambridge, 2000.

\bibitem{lectures} G. Lazarides, Lect. Notes
Phys. 592 (2002) 351 [hep-ph/0111328];
%\lnp{592}{2002}{351}
%%CITATION = HEP-PH 0111328;%%
hep-ph/0204294.
%%CITATION = HEP-PH 0204294;%%

\bibitem{curv1}
S. Mollerach, \prd{42}{1990}{313};
%%CITATION = PHRVA,D42,313;%%
A.D. Linde and V. Mukhanov, ibid. 56 (1997)
535.
%\ibid{56}{1997}{535}.
%%CITATION = ASTRO-PH 9610219;%%

\bibitem{curv2}
D.H. Lyth and D. Wands, \plb{524}{2002}{5};
%%CITATION = HEP-PH 0110002;%%
T. Moroi and T. Takahashi, \ibid{522}{2001}{215};
539 (2002) 303(E).
%%CITATION = HEP-PH 0110096;%%

\bibitem{curv3}
D.H. Lyth, C. Ungarelli and D. Wands,
\prd{67}{2003}{023503}.
%%CITATION = ASTRO-PH 0208055;%%

\bibitem{dimo}
K. Dimopoulos and D.H. Lyth, Phys. Rev. D
69 (2004) 123509.
%\prd{69}{2004}{123509}.
%%CITATION = HEP-PH 0209180;%%

\bibitem{wmap1}
C.L. Bennett et al., Astrophys. J. Suppl.
148 (2003) 1;
%\apjs{148}{2003}{1};
%%CITATION = ASTRO-PH 0302207;%%
G. Hinshaw et al., ibid. 148 (2003) 135;
%\ibid{148}{2003}{135};
%%CITATION = ASTRO-PH 0302217;%%
A. Kogut et al., \ibid{148}{2003}{161};
%%CITATION = ASTRO-PH 0302213;%%
D.N. Spergel et al., \ibid{148}{2003}{175}.
%%CITATION = ASTRO-PH 0302209;%%

\bibitem{wmap2}
H.V. Peiris et al., Astrophys. J. Suppl.
148 (2003) 213.
%\apjs{148}{2003}{213}.
%%CITATION = ASTRO-PH 0302225;%%

\bibitem{iso1}
C. Gordon and A. Lewis, Phys. Rev. D 67 (2003)
123513;
%\prd{67}{2003}{123513};
%%CITATION = ASTRO-PH 0212248;%%
P. Crotty, J. Garc{\'\i}a-Bellido, J. Lesgourgues
and A. Riazuelo, \prl{91}{2003}{171301};
%%CITATION = ASTRO-PH 0306286;%%
M. Bucher, J. Dunkley, P.G. Ferreira, K. Moodley
and C. Skordis, \ibid{93}{2004}{081301};
%%CITATION = ASTRO-PH 0401417;%%
K. Moodley, M. Bucher, J. Dunkley, P.G. Ferreira
and C. Skordis, \prd{70}{2004}{103520}.
%%CITATION = ASTRO-PH 0407304;%%

\bibitem{gm}
C. Gordon and K.A. Malik, Phys. Rev. D 69
(2004) 063508;
%%CITATION = ASTRO-PH 0311102;%%
M. Beltr\'{a}n, J. Garc{\'\i}a-Bellido, J.
Lesgourgues and A. Riazuelo, ibid. 70
(2004) 103530.
%%CITATION = ASTRO-PH 0409326;%%

\bibitem{iso}
G. Lazarides, R. Ruiz de Austri and R. Trotta,
\prd{70}{2004}{123527}.
%%CITATION = HEP-PH 0409335;%%

\bibitem{hier}
G. Lazarides and N.D. Vlachos,
%\plb{459}{1999}{482}.
Phys. Lett. B 459 (1999) 482;
%%CITATION = HEP-PH 9903511;%%
G. lazarides, hep-ph/9905450.
%%CITATION = HEP-PH 9905450;%%

\bibitem{lyth}
E.J. Copeland, A.R. Liddle, D.H. Lyth,
E.D. Stewart and D. Wands,
Phys. Rev. D 49 (1994) 6410.
%\prd{49}{1994}{6410}.
%%CITATION = ASTRO-PH 9401011;%%

\bibitem{dss} G. Dvali, Q. Shafi and
R. Schaefer, \prl{73}{1994}{1886}.
%%CITATION = HEP-PH 9406319;%%

\bibitem{dllr1}
K. Dimopoulos, G. Lazarides, D. Lyth and R.
Ruiz de Austri, \jhep{05}{2003}{057}.
%%CITATION = HEP-PH 0303154;%%

\bibitem{rsym} G. Lazarides and Q. Shafi,
Phys. Rev. D 58 (1998) 071702.
%\prd{58}{1998}{071702}.
%%CITATION = HEP-PH 9803397;%%

\bibitem{hybrid} A.D. Linde,
\plb{259}{1991}{38};
%%CITATION = PHLTA,B259,38;%%
\prd{49}{1994}{748}.
%%CITATION = ASTRO-PH 9307002;%%

\bibitem{kn} J.E. Kim and H.P. Nilles,
\plb{138}{1984}{150}.
%%CITATION = PHLTA,B138,150;%%

\bibitem{crisis}
M. Dine, W. Fischler and D. Nemeschansky,
\plb{136}{1984}{169};
%%CITATION = PHLTA,B136,169;%%
G.D. Coughlan, R. Holman, P. Ramond and
G.G. Ross, \ibid{140}{1984}{44}.
%%CITATION = PHLTA,B140,44;%%

\bibitem{drt95}
M. Dine, L. Randall and S. Thomas,
\prl{75}{1995}{398};
%%CITATION = HEP-PH 9503303;%%
Nucl. Phys. B 458 (1996) 291.
%\npb{458}{1996}{291}.
%%CITATION = HEP-PH 9507453;%%

\bibitem{noscale}
E.D. Stewart, \prd{51}{1995}{6847};
%%CITATION = HEP-PH 9405389;%%
M.K. Gaillard, H. Murayama and K.A. Olive,
\plb{355}{1995}{71};
%%CITATION = HEP-PH 9504307;%%
M.K. Gaillard, D.H. Lyth and H. Murayama,
\prd{58}{1998}{123505};
%%CITATION = HEP-TH 9806157;%%
C. Panagiotakopoulos, \plb{459}{1999}{473};
%%CITATION = HEP-PH 9904284;%%
R. Jeannerot, S. Khalil and G. Lazarides,
\jhep{07}{2002}{069}.
%%CITATION = HEP-PH 0207244;%%

\bibitem{ekn}
J.R. Ellis, J.E. Kim and D.V. Nanopoulos,
\plb{145}{1984}{181};
%%CITATION = PHLTA,B145,181;%%
J.R. Ellis, D.V. Nanopoulos and S. Sarkar,
\npb{259}{1985}{175};
%%CITATION = NUPHA,B259,175;%%
J.R. Ellis, G.B. Gelmini, J.L. Lopez,
D.V. Nanopoulos and S. Sarkar,
\ibid{373}{1992}{399}.
%%CITATION = NUPHA,B373,399;%%

\bibitem{reheat}
R.J. Scherrer and M.S. Turner, Phys. Rev. D
31 (1985) 681 .
%%CITATION = PHRVA,D31,681;%%

\bibitem{communication}
K. Dimopoulos and D.H. Lyth, private
communication.

\bibitem{notari}
K. Dimopoulos, D.H. Lyth, A. Notari and A. Riotto,
\jhep{07}{2003}{053}.
%%CITATION = HEP-PH 0304050;%%

\bibitem{evap}
R. Allahverdi, B.A. Campbell and J.R. Ellis,
\npb{579}{2000}{355};
%%CITATION = HEP-PH 0001122;%%
A. Anisimov and M. Dine,
ibid. 619 (2001) 729.
%%CITATION = HEP-PH 0008058;%%

\bibitem{lepto}
M. Fukugita and T. Yanagida, Phys. Lett. B
174 (1986) 45.
%\plb{174}{1986}{45}.
%%CITATION = PHLTA,B174,45;%%

\bibitem{leptoinf}
G. Lazarides and Q. Shafi, \plb{258}{1991}{305};
%%CITATION = PHLTA,B258,305;%%
G. Lazarides, R.K. Schaefer and Q. Shafi,
\prd{56}{1997}{1324}.
%%CITATION = HEP-PH 9608256;%%

\bibitem{km}
M. Kawasaki and T. Moroi, Prog. Theor. Phys.
93 (1995) 879.
%\ptp{93}{1995}{879}.
%%CITATION = HEP-PH 9403364;%%

\bibitem{axion}
M.S. Turner, \prd{33}{1986}{889}.
%%CITATION = PHRVA,D33,889;%%

\bibitem{pivot}
E.J. Copeland, I.J. Grivell and A.R. Liddle,
astro-ph/ 9712028.
%%CITATION = ASTRO-PH 9712028;%%

\bibitem{iso3}
L. Amendola, C. Gordon, D. Wands and M. Sasaki,
\prl{88}{2002}{211302}.
%%CITATION = ASTRO-PH 0107089;%%

\bibitem{thesis}
R. Trotta, astro-ph/0410115.
%%CITATION = ASTRO-PH 0410115;%%

\bibitem{cobe10}
M. Tegmark and A.J.S. Hamilton,
astro-ph/9702019.
%%CITATION = ASTRO-PH 9702019;%%

\bibitem{hst}
W.L. Freedman et al.,
Astrophys. J. 553 (2001) 47.
%\apj{553}{2001}{47}.
%%CITATION = ASTRO-PH 0012376;%%

\bibitem{cosmomc}
\textsf{http://cosmologist.info/cosmomc/}.

\bibitem{lewis}
A. Lewis and S. Bridle,
Phys. Rev. D 66 (2002) 103511.
%\prd{66}{2002}{103511}.
%%CITATION = ASTRO-PH 0205436;%%

\bibitem{Valiviita}
J. V\"{a}liviita and V. Muhonen,
Phys. Rev. Lett. 91 (2003) 131302.
%\prl{91}{2003}{131302}.
%%CITATION = ASTRO-PH 0304175;%%

\bibitem{Verde}
L. Verde et al.,
Astrophys. J. Suppl. 148 (2003) 195.
%\apjs{148}{2003}{195}.
%%CITATION = ASTRO-PH 0302218;%%

\bibitem{cbi}
T.J. Pearson et al.,
Astrophys. J. 591 (2003) 556.
%\apj{591}{2003}{556}.
%%CITATION = ASTRO-PH 0205388;%%

\bibitem{acbar}
J.H. Goldstein et al.,
Astrophys. J. 599 (2003) 773;
%\apj{599}{2003}{773};
%%CITATION = ASTRO-PH 0212517;%%
C.-l. Kuo et al., \ibid{600}{2004}{32};
%%CITATION = ASTRO-PH 0212289;%%
\textsf{http://cosmologist.info/ACBAR}.

\bibitem{Burles}
S. Burles, K.M. Nollett and M.S. Turner,
\prd{63}{2001}{063512}.
%%CITATION = ASTRO-PH 0008495;%%

\end{thebibliography}
\end{document}